\documentclass[prl,twocolumn,superscriptaddress,amsfonts,amssymb]{revtex4}
\usepackage{graphicx}
\usepackage{dcolumn}
\usepackage{xcolor}
\usepackage{color} 
\usepackage{bm}
\usepackage{times}
\usepackage{color, soul}

\begin{document}

\newcommand{\de }{$^{\circ}$}
\newcommand{\mr}[1]{{\textcolor{red} {#1}}}
\newcommand{\mi}[1]{{\textcolor{blue} {#1}}}
\newcommand{\mj}[1]{{\textcolor{orange} {#1}}}
\newcommand{\JL}[1]{\textcolor{red}{{\bf JL: #1 }}} 
\newcommand{\IPZ}[1]{\textcolor{purple}{{\bf IPZ: #1 }}}

\title{Tunable energy and mass renormalization from homothetic Quantum dot arrays}

\author{Ignacio Piquero-Zulaica* }
\affiliation{Centro de F\'{\i}sica de Materiales CSIC/UPV-EHU-Materials Physics Center, Manuel Lardizabal 5, E-20018 San Sebasti\'an, Spain}

\author{Jun Li}
\affiliation{Zernike Institute for Advanced Materials, University of Groningen, Nijenborgh 4, 9747 AG Groningen, The Netherlands }

\author{Zakaria M. Abd El-Fattah}
\affiliation{Physics Department, Faculty of Science, Al-Azhar University, Nasr City E-11884 Cairo, Egypt}
\affiliation{ICFO-Institut de Ciencies Fotoniques, The Barcelona Institute of Science and Technology, 08860 Castelldefels,Barcelona, Spain}

\author{Leonid Solianyk}
\affiliation{Zernike Institute for Advanced Materials, University of Groningen, Nijenborgh 4, 9747 AG Groningen, The Netherlands }

\author{Iker Gallardo}
\affiliation{Centro de F\'{\i}sica de Materiales CSIC/UPV-EHU-Materials Physics Center, Manuel Lardizabal 5, E-20018 San Sebasti\'an, Spain}

\author{Leticia Monjas}
\affiliation{Institute for Chemistry, University of Groningen, Nijenborgh 7, 9747 AG Groningen, the Netherlands}

\author{Anna K. H. Hirsch}
\affiliation{Stratingh Institute for Chemistry, University of Groningen, Nijenborgh 7, 9747 AG Groningen, The Netherlands.}
\affiliation{Helmholtz Institute for Pharmaceutical Research Saarland (HIPS) - Helmholtz Centre for Infection Research (HZI), Department of Drug Design and Optimization andDepartment of Pharmacy, Saarland University, Campus Building E8.1, 66123  Saarbr{\"{u}}cken, Germany}

\author{Andres Arnau}
\affiliation{Centro de F\'{\i}sica de Materiales CSIC/UPV-EHU-Materials Physics Center, Manuel Lardizabal 5, E-20018 San Sebasti\'an, Spain}
\affiliation{Donostia International Physics Center, Paseo Manuel Lardizabal 4, E-20018 Donostia-San Sebasti\'an, Spain}
\affiliation{Dpto. de F\'{\i}sica de Materiales, Universidad del Pa\'{\i}s Vasco, E-20018 San Sebasti\'an, Spain }

\author{J. Enrique Ortega}
\affiliation{Centro de F\'{\i}sica de Materiales CSIC/UPV-EHU-Materials Physics Center, Manuel Lardizabal 5, E-20018 San Sebasti\'an, Spain}
\affiliation{Donostia International Physics Center, Paseo Manuel Lardizabal 4, E-20018 Donostia-San Sebasti\'an, Spain}
\affiliation{Universidad del Pa\'{\i}s Vasco, Dpto. F\'{\i}sica Aplicada I,  E-20018 San Sebasti\'an, Spain}

\author{Meike St\"{o}hr* }
\affiliation{Zernike Institute for Advanced Materials, University of Groningen, Nijenborgh 4, 9747 AG Groningen, The Netherlands }

\author{Jorge Lobo-Checa* }
\affiliation{Instituto de Ciencia de Materiales de Arag\'on (ICMA), CSIC-Universidad de Zaragoza, E-50009 Zaragoza, Spain}
\affiliation {Departamento de F\'{\i}sica de la Materia Condensada, Universidad de Zaragoza, E-50009 Zaragoza, Spain}

\date{\today}

\begin{abstract}
 Quantum dot arrays in the form of molecular nanoporous networks are renown for modifying the electronic surface properties through quantum confinement. Here we show that, compared to the pristine surface state, the fundamental energy of the confined states can exhibit downward shifts accompanied by a lowering of the effective masses simultaneous to the appearance of tiny gaps at the Brillouin zone boundaries. We observed these effects by angle resolved photoemission for two self-assembled homothetic (scalable) Co-coordinated metal-organic networks. Complementary scanning tunneling spectroscopy measurements confirmed these findings. Electron plane wave expansion simulations and density functional theory calculations provide insight into the nature of this phenomenon, which we assign to metal-organic overlayer-substrate interactions in the form of adatom-substrate hybridization. 
The absence to date of the experimental band structure resulting from single adatom metal-coordinated nanoporous networks has precluded the observation of the significant surface state renormalization reported here, which we infer are general of low interacting and well-defined adatom arrays.
 
* Corresponding Authors
\end{abstract}


\maketitle

Over the last decades, the concepts of supramolecular chemistry have been successfully transferred to the construction of two-dimensional (2D) self-assembled molecular arrangements on metallic surfaces \cite{Lehn1995,Atwood1996, Stepanow2008a, Muller2016}.
By selecting the proper tectons (molecular constituents and, if required, metal linkers) and depositing them onto selected substrates, long-range ordered, regular and robust nanoporous networks have been achieved, ranging from hydrogen- \cite{Pawin2006} or halogen-bonded \cite{Piquero2017}, to metal-organic structures \cite{Matena2014,NianLin2016}. Such regular structures stand out as ideal templates for nanopatterning organic and inorganic adsorbates by selective adsorption  \cite{Stohr2007,Bartels2010,Wyrick2011, Nowakowska2015, NianLin2015, Pivetta2013}. Nanoporous networks, also referred to as quantum dot (QD) arrays since they can confine surface state (SS) electrons, provide a vast playground for studying and engineering the electronic properties of new and exotic 2D materials. Indeed, metal-organic networks are known to show novel magnetic properties \cite{Stepanow2013, Umbach2012}, catalytic effects \cite{Kern2015}, oxidation states \cite{Gottfried2012}, exotic tesellation \cite{Urgel2016,NianLin2017,Florian2018} and bear the prospect of exhibiting topological electronic bands \cite{Zhao2015, Zhang2016}.\\
\indent 
The dominant electronic signature around the Fermi level due to the presence of nanoporous networks comes from the substrate's surface state electrons, which scatter at the molecular array and become confined within individual nanopores \cite{Lobo2009,Florian2011}. The tunability of the confined electronic state has so far been achieved by varying the pore dimensions, i.e. the QD size \cite{Florian2011,Wackerlin2014}. However, since the confining potential barriers are not infinite, these QDs are not independent but coupled, as has been shown by angle resolved photoemission (ARPES) through the existence of new dispersive electronic bands \cite{Lobo2009}, as well as by Fourier-transform scanning tunneling spectroscopy (FT-STS) data~\cite{NianLin2013b}. 
These QD array bands can be modified through the condensation of guest atoms  \cite{Nowakowska2016} or by changing the barrier width \cite{Piquero2017}. The standard fingerprints, whenever confinement of two-dimensional electron gases (2DEGs) occurs on noble metal surfaces, are in the form of an energy shift towards the Fermi level of its fundamental energy, an increase of the effective mass, and the appearance of energy gaps at the surface Brillouin zone (BZ) boundaries \cite{Lobo2009,Piquero2017,Mugarza2006,Bendounan2006}.\\
\indent
In this work we show for two homothetic (scalable) metal-organic nanoporous networks (MONN) grown on Au(111) the first experimental evidence of energy downshifts and reduced effective masses compared to the pristine SS, simultaneous to the opening of zone boundary gaps that suggest electron confinement within the nanocavities. More specifically, these effects are gradual, i.e. they depend on the network dimensions. 
The interaction between the Au substrate and the MONN are at the base of these unexpected phenomena and it is not a consequence of the quantum confinement. \\
\begin{figure*}
\begin{center}
  \includegraphics[width=0.9\textwidth,clip]{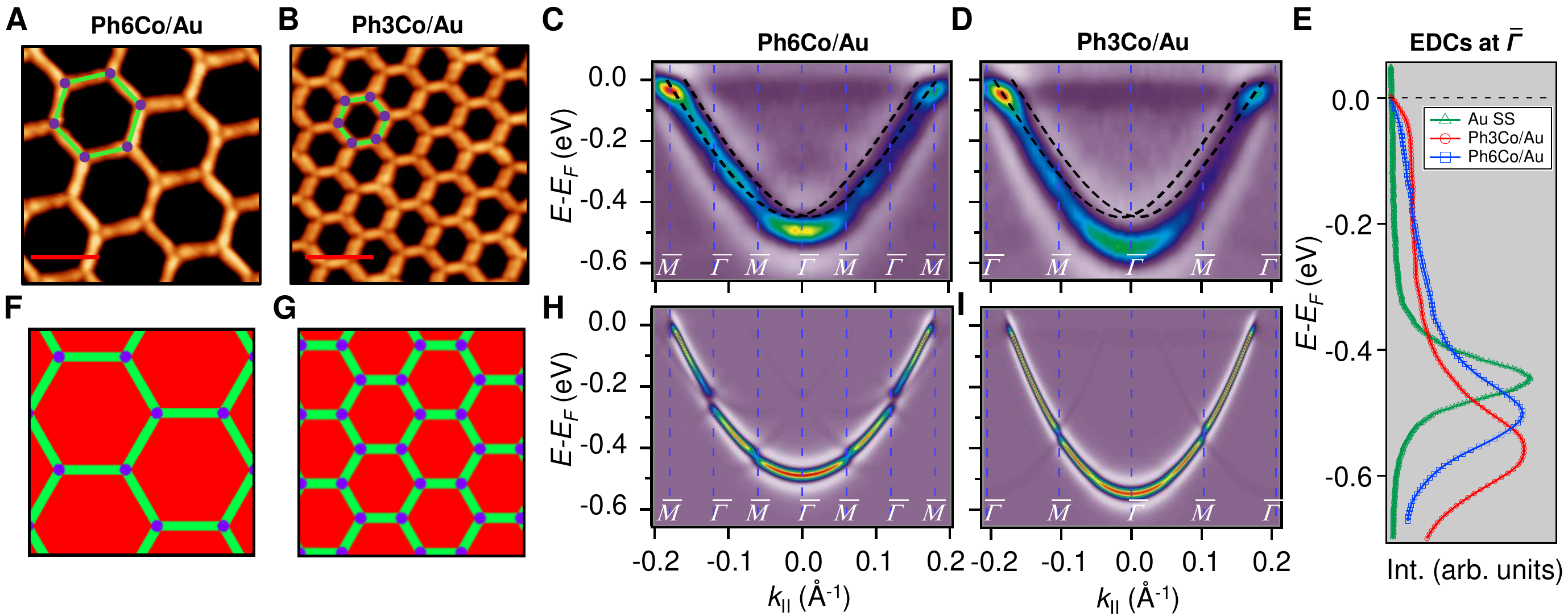}
 \end{center}
\vspace*{-8mm}
\caption[] { STM topographies of the single domain Co-coordinated hexagonal QD arrays using (A) dicarbonitrile-sexyphenyl (Ph6) and (B) dicarbonitrile-terphenyl (Ph3). Scale bar in red corresponds to 5 nm. (C, D) Second derivative of the spectral density obtained by ARPES at $150$~K along the $\overline{\Gamma \rm{M}}$ high-symmetry direction for both Ph6Co and Ph3Co nanoporous networks. The band structure exhibits downward shifts of the band bottom and gap openings at the superstructure symmetry points compared to the pristine Au(111) Shockley state (black dotted lines). (E) Energy distribution curves (EDCs) at normal emission ($\overline{\Gamma}$ point) for pristine Au(111) (green), Ph6Co (blue) and Ph3Co (red). A gradual downshift of the fundamental energy as the pore size is reduced ($\Delta E_{Ph6Co} = 40$~meV and $\Delta E_{Ph3Co} = 100$~meV with respect to the Au SS) is found. 
(F, G) 2D potential geometry used for the EPWE modelization, where green stands for the molecular repulsive potentials, purple for slightly repulsive Co regions and red for cavity regions with zero potential. (H,  I) Band structure along $\overline{\Gamma \rm{M}}$ direction of the overlayers simulated by EPWE based on the previous geometry. Matching the experimental ARPES data (gap openings and band-bottom shifts) requires a significant modification of the 2DEG energy reference (see text for details).
}
\label{figure1}
\end{figure*}

The studied scalable Co-coordinated networks were grown on Au(111) from two related dicarbonitrile-polyphenyl derivates. 
Specifically, we used dicarbonitrile-terphenyl (Ph3) and dicarbonitrile-sexyphenyl (Ph6) molecules and Co atoms in a 3:2 stoichiometry to fabricate the MONN. These tectons were sequentially evaporated (molecules first, then Co) onto Au(111) followed by a mild annealing to 400~K. That resulted in two scalable, periodic, long-range order and practically defect free QD arrays [shown in Figure \ref {figure1}(A-B)] and named hereafter Ph6Co and Ph3Co. In agreement with previous work \cite{Florian2011}, the networks exhibit sixfold symmetry with unit cell vectors of $3.53$~nm (for Ph3Co) and $5.78$~nm (for Ph6Co) along the [11$\overline{2}$] direction and enclose pore areas of $8$~nm$^{2}$ and $24$~nm$^{2}$, respectively. Note that the interaction of both networks with the substrate is assumed to be rather weak since the herringbone reconstruction is neither lifted nor modified in its periodicity \cite{Supplementary,Oses2009}.
We experimentally probed these networks with ARPES [Helium I source ($h\nu$=$21.2$~eV) at $150$~K] and scanning tunneling microscopy/spectroscopy (STM/STS) at 5 K to obtain both spatially averaged and spatially highly resolved information  \cite{Supplementary}. The experimental data are complemented by Electron plane wave expansion (EPWE) simulations and Density Functional Theory (DFT) calculations \cite{Supplementary}.\\
\indent
The 2DEG onset of Ph6Co and Ph3Co networks formed on Au(111) is reliably determined by ARPES and only approximately by STS \cite{Piquero2017}. Moreover, ARPES --in contrast to STS-- can resolve the QD array band structure from the MONN. However, this can be exceedingly challenging because the networks must be extended, almost defect-free and completely covering the probed surface (in absence of other coexistent molecular phases) \cite{Lobo2009, Piquero2017}. To achieve these conditions we evaporated the molecules and  Co adatoms in  orthogonal shallow gradient depositions on the Au(111) substrate, thereby ensuring the existence of an area with optimal coverage and the exact 3:2 stoichiometry \cite{Piquero2016}. 
Figure \ref {figure1}(C, D) shows the second derivative of the ARPES spectral density from Ph6Co and Ph3Co along the $\overline{\Gamma \rm{M}}$ high symmetry direction.  We observed a gradual downshift of the fundamental energy ($\overline{\Gamma}$ point) towards higher binding energies as the pore size is reduced, which can be quantified from the normal emission energy distribution curves (EDCs) [cf. Fig. \ref {figure1}(E) and Table \ref {table1}]. Note that this clearly goes in the opposite direction to the energy shift expected from conventional lateral confinement systems. Simultaneously to this downshift, we observed a reduction of the effective mass (see table \ref {table1}), resembling a Fermi wave-vector pinning \cite{Supplementary}. The partial confinement of the substrate's 2DEG is inferred from the presence of small gaps (observed as slight intensity variations) at the symmetry points, which denotes weak scattering from the network barriers. Note that the absence of spin-orbit splitting in our data for Ph3Co and Ph6Co does not rule out this effect, as it could be masked by ARPES lineshape intrinsic broadening \cite{Supplementary}.\\
\begin{table}
	\begin{center}
		\begin{tabular}{c|c|c||c|c|}
					 & $E^{\overline{\Gamma}}_B$ (eV)  & $m^*/m_0$  &  $E^{Ref, \overline{\Gamma}}_{EPWE}$ (eV) & $m^{*,Ref}_{EPWE} / m_0$\\
			\hline 
			Au(111) & 0.45 & 0.255 & 0.45 & 0.26\\
			\hline
			Ph6Co & 0.49 & 0.24 & 0.52 & 0.24 \\
			\hline
			Ph3Co & 0.55 & 0.22 &  0.59 & 0.21 \\			
			\hline
		\end{tabular}
	\end{center}
	\caption[]{ARPES experimental binding energies at $\overline{\Gamma}$ and effective masses  (columns $E^{\overline{\Gamma}}_B$ and $m^*/m_0$)  for the substrate and the two networks. The corresponding 2DEG references (fundamental energy and effective masses) required for matching ARPES with the EPWE simulations are indicated in the last two columns: $E^{Ref, \overline{\Gamma}}_{EPWE}$ and $m^{*,Ref}_{EPWE} / m_0$.}
\label{table1}
\end{table}
\indent

To unravel the potential energy landscapes generated by the molecular networks and their confining properties, we performed EPWE simulations. Such a semi-empirical model has been successfully used for similar systems~\cite{Barth2009, Florian2011, Piquero2017}. The geometry of both systems for the simulations were defined following topographic STM images [see Figures \ref {figure1}(F, G)]. Assuming repulsive scattering potential sites for molecules ($V_{mol} = 250$~meV) and Co atoms ($V_{Co} = 50$~meV), the experimental data were correctly reproduced. In particular, the ARPES energy gaps ($\sim 25$~meV for Ph6Co and $\sim 30$~meV for Ph3Co  at $\overline{\rm{M}}$) reflect the weak scattering strength of the networks [Figures \ref {figure1}(H, I)]. However, such repulsive scattering is known to shift the 2DEG fundamental energy (at $\overline{\Gamma}$) upwards, opposite to what is observed here. In this way, the ARPES dispersions can only be matched by EPWE when adopting higher binding energy references and smaller effective masses than the pristine Au(111) SS (see Table \ref{table1}). In other words, using the original dispersion of the Au(111) SS as scattering reference cannot correctly reproduce the experimental data.\\
\begin{figure}
\begin{center}
  \includegraphics[width=0.50\textwidth,clip]{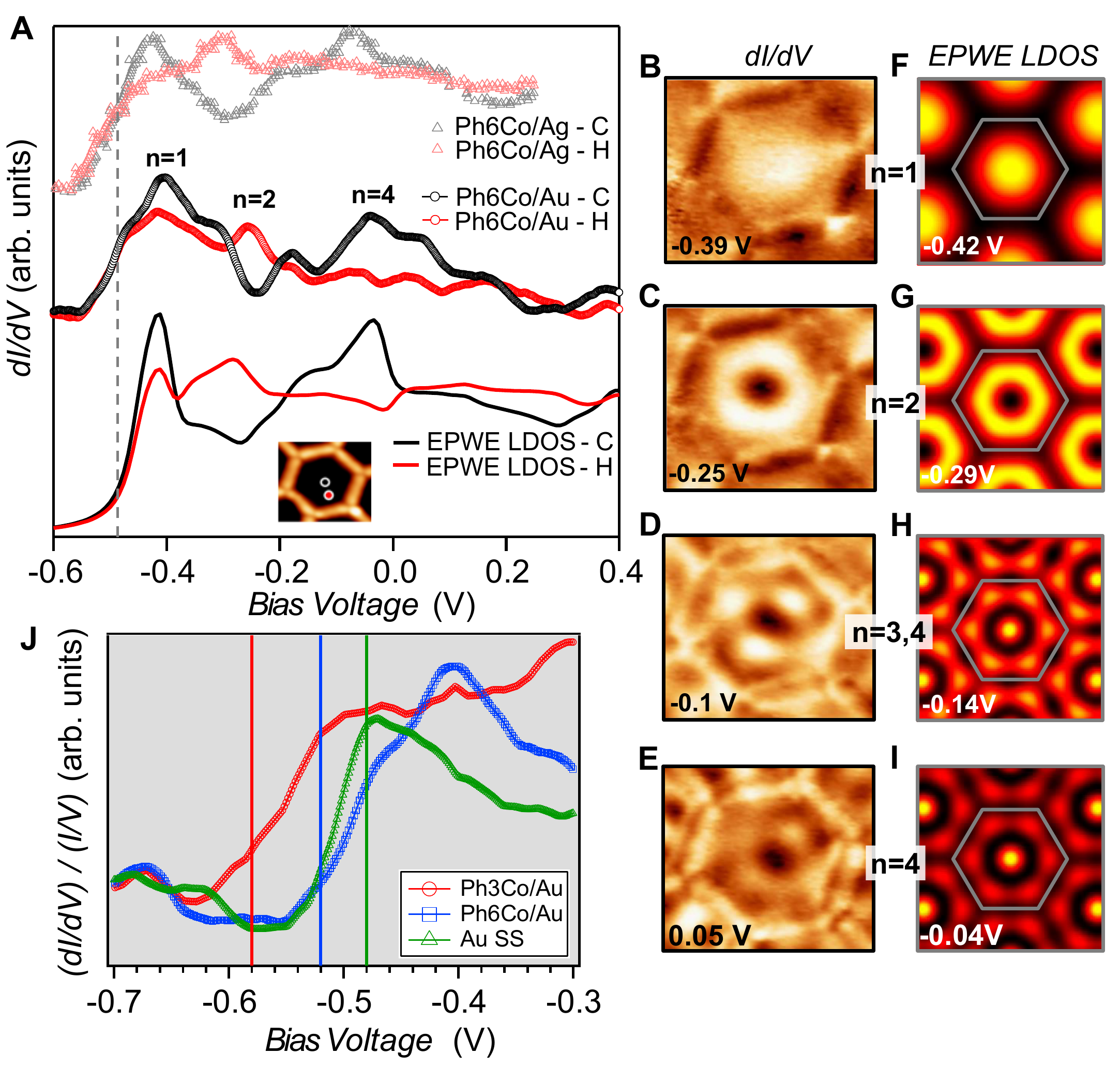}
 \end{center}
\vspace*{-8 mm}
\caption[]{ Local confinement and renormalization effects observed by STM/STS. (A) $dI/dV$ spectra at the pore center (black) and halfway (red) for three Ph6Co datasets: Experimental curves of Ph6Co on Au(111) (middle), corresponding EPWE conductance simulation using the ARPES parameters (bottom), and experimental spectra of Ph6Co on Ag(111) adapted from Ref.~\onlinecite{Florian2011} and normalized (see the text) to the Au(111) 2DEG (top). The spectra are made up of the characteristic confined state resonances that alternate depending on the wavefunction spatial distribution, i.e., $n=1$ and $n= 4$ peak at the pore center  and $n=2$ at halfway~\cite{Berndt1999,Pivetta2013}. (B to E) Experimental $dI/dV$ maps reproducing standing wave patterns of the different energy levels $n$ showing excellent agreement with the EPWE simulated ones at similar energies (F to I). The observed deviations  for the higher energy conductance maps can be assigned to weak potential alterations stemming from the underlying herringbone reconstruction \cite{Supplementary}, which are not considered in the simulations (see text for details). (J) Zoom-in onto the experimental $dI/dV$ onset for the pristine Au(111) SS (green) and Ph6Co (blue) and Ph3Co (red) networks probed at the center of the pores. A downshift of the onset energy is observed that agrees with the ARPES fundamental energy positions (temperature corrected by 30 meV) that are indicated in the panel by vertical thin lines.
}
\label{figure2}
\end{figure}
\indent
Such an unexpected scenario questions the confining capabilities of these MONN. Using STS, we could verify that these networks do confine the Au SS, similarly to the ones generated onto Ag(111) by the same family of molecules \cite{Florian2011}. In panel (A) (middle) of Figure \ref{figure2} we present the Ph6Co STS data acquired at two different positions. The conductance spectra together with the $dI/dV$ maps taken at different voltages [Figure \ref {figure2}(B to E)] exhibit clear confinement resonances within the pores \cite{Barth2009,Lobo2009,Florian2011,Pivetta2013,Nowakowska2016,Piquero2017}. Such electron localization mirrors the one observed for the same network  on Ag(111) \cite{Florian2011}.
In order to directly compare them we adapt the $dI/dV$ spectra of ref.~\onlinecite{Florian2011}  by normalizing the energy axis by the ratio of the respective effective masses ($m^{*,Ph6Co}_{Ag} / m^{*,Ph6Co}_{Au} = 0.41/0.24$) and shifting the onset of the Ag SS to the one of Au ($-485$~meV at 5 K). The agreement (lineshape and peak energies) between the two datasets is quite reasonable [cf. middle and top of Figure \ref {figure2}(A)], demonstrating that the confinement properties of Ph6Co are similar for the two substrates.\\
\indent

We can now address the 2DEG energy downshift with respect to the Au SS upon network formation using local techniques. The overall  $dI/dV$ lineshapes at the pore center exhibit broad peak widths (reflecting the ARPES bandwidth) and are quite asymmetric (with maxima being displaced towards higher energy) \cite {Piquero2017}. Such spectral asymmetry for $n=1$ at the pore center can be understood from a band structure perspective: the reduced onset contribution relates to electrons spreading out over the surface given their longer wavelength ($\lambda = 2\pi/k$) at the band bottom ($k\sim 0$ around $\overline{\Gamma}$). Contrarily, the STS is maximized at higher energies (close to the $\overline{M}$ point) since the electrons have much shorter wavelengths, thereby becoming much more sensitive to the network barriers and prone to be trapped within the pores. 
Figure \ref {figure2}(J), shows the STS spectra of the two networks at the pore center compared to the Au(111) SS. For Ph3Co the onset is clearly shifted away from the Au SS onset, whereas for Ph6Co it is similar but still slightly displaced. This is also the case for these networks on Ag(111) (see S.I. \cite{Supplementary}). For the Ph6Co, we simulated the STS point spectra and conductance maps [Figure \ref {figure2}(A, F-I)] using the same scattering parameters and effective mass reduction as described above for the ARPES electron bands. While the experimental and simulated STS spectra match reasonably well, we observe slight discrepancies for the conductance maps taken at higher energies. This can be ascribed to weak potential variations introduced by the reconstruction of the underlying substrate. Indeed the Ph6Co unit cell is large enough to host  both fcc and hcp regions within a single pore (cf. S.I.  \cite{Supplementary}), which was not accounted for by the EPWE simulations.\\
\indent
In essence, the STS shifts qualitatively agree with the ARPES results, as observed in Fig. \ref {figure2}(J) (vertical lines), supporting a change of the 2DEG reference upon network presence on the surface. We believe that a subtle downward energy shift, as it is the case of Ph3Co and Ph6Co, may also exist for other MONN \cite {NianLin2013b, Florian2011, Pivetta2013}. However, since complementary photoemission experiments are required for observing this 2DEG onset reference for the QD array, the present effect has not been reported up to now.\\
\begin{figure}
\begin{center}
  \includegraphics[width=0.4\textwidth,clip]{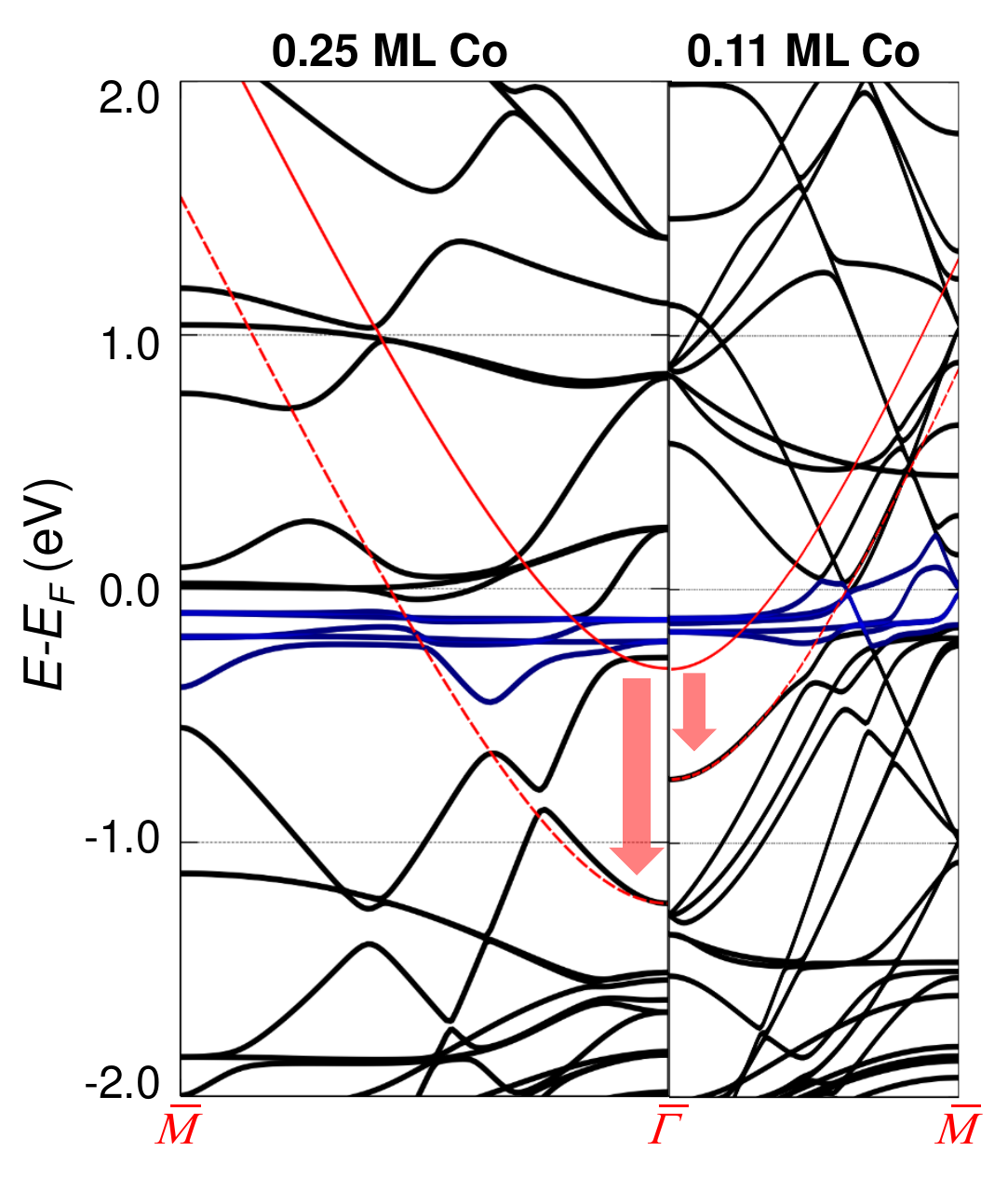}
 \end{center}
\vspace*{-7 mm}
\caption[]{Visualization of the Au(111) surface state (continuous red curve for the pristine case)  downward energy shift  at two different Co coverages. The vertical arrows show the calculated shift close to $\overline{\Gamma}$ and the red dotted lines are a guide to the eye to follow the altered SS. The left panel corresponds to 0.25 ML of Co and the right panel to 0.11 ML, as obtained using a 2x2 and a 3x3 surface unit cell, respectively. The different supercells introduce an evident difference in the folding of the Au bands. The blue curves near the Fermi level correspond to Co d-bands. The coupling between the Co d-bands and folded bulk-bands with the Au(111) surface state pushes it downwards in energy, the shift being larger at higher Co coverages. 
}
\label{fig3}
\end{figure}
\indent

Different factors might be responsible for these counter-intuitive downward energy shifts of the confined states with respect to the Au SS. This effect can be attributed to the network-substrate interactions in the form of charge transfer (doping effects) or hybridization effects of the metal adsorbates that renormalize the  2DEG that is modulated by the network potential landscape. 
As the shift is gradual, being larger for Ph3Co than for Ph6Co, and the networks are homothetic, it could be induced by charge transfer from the Co adatoms [full surface coverage of Ph3Co/Ph6Co corresponds to 0.015/0.005 monolayers (ML) of Co] to the Au SS, similar to the downshift induced by alkali metals~\cite {Memmel1996}. However, the fact that $m^*$ decreases and the Fermi wave-vector ($k_{F}$) is practically pinned suggests the conservation of the 2DEG electron occupancy (the electron density $n=\frac{k_{F}^{2}}{2\pi}$) \cite {Reinert2006, Liu2006}. Therefore, the Au SS shift is not driven by electron charge transfer from the Co atom to the Au surface. \\
\indent
The downward shift of the Shockley state may reflect the Co interaction with the Au substrate, that is, the local Co/Au hybridization \cite{Schlickum2007}. In this way, we explore the weak Co-Au hybridization by means of DFT calculations of Co atom arrays onto a non-reconstructed Au(111) surface. Figure~\ref{fig3} shows the calculated band structure from two selected supercells:  2x2 (0.25 ML) on the left and  3x3 (0.11 ML) on the right. These superstructures introduce an evident difference in the folding of the Au bands (in black), but more importantly, a clear downshift of the pristine Au SS (red arrow). We find that the magnitude of the downshift is directly related to the  amount of isolated Co adatoms on the surface (see S.I. \cite{Supplementary}). The actual Co coverage within the networks is much lower (by about an order of magnitude), so the expected shift obtained by simple extrapolation to the corresponding Co coverage (of the order of 50 meV) is comparable to the experimental observations (cf. the SI \cite{Supplementary}). Although geometrical variations (vertical displacements) of the overlayers \cite {NianLin2016b, Matena2014, Xu2016} that could affect the SS reference cannot be completely discarded, the hybridization (coupling) of the Co d-bands (shown in blue in  Fig. \ref{fig3}) and folded bulk-bands with the Au(111) surface state convincingly explains the observed SS renormalization effect \cite{Liu2006,Footnote1}. \\
\indent
This effect turns out to be more general than initially expected. First, because it is likewise  experimentally observed for this family of MONN grown onto a different noble metal substrate  (cf. the SI \cite{Supplementary}), and second since additional DFT calculations for homoatomic arrays [Cu/Cu(111) and Au/Au(111)] exhibit the same effect (cf. SI \cite{Supplementary}). We deduce that this holds for (homo- and hetero-) atomic arrays formed onto noble metal substrates whenever the hybridization is not strong (physisorption cases), such that the SS character is maintained. This commonly applies to MONN since the molecules slightly pull the adatoms away from the surface \cite {NianLin2016b, Matena2014, Xu2016}, effectively reducing the interaction.\\

\indent

In summary, ARPES and STS results reveal a gradual energy and mass renormalization of the Au(111) SS upon formation of two homothetic Co coordinated metal-organic networks. EPWE simulations only agree with the experimental data after the 2DEG reference is shifted to higher binding energies. Notably this downshift is gradual with decreasing pore size and is observable in spite of the confining attributes of the nanocavities (that upshifts the states). Our EPWE simulations can satisfactorily match our experimental data using repulsive potentials for both molecules and Co atoms. 
Overlayer-substrate interactions must be responsible for such counterintuitive effects upon the Au SS reference. Hybridization between the Co adatoms and the folded susbtrate bands with the Au SS appear as the most plausible cause, as deduced from DFT calculations. We predict that other MONNs grown on noble metal surfaces should show such subtle counterintuitive 2DEG energy renormalization whenever the SS character is preserved, i.e. for weak coupling cases.\\
\indent
\\

We acknowledge Prof. J. Garc\'ia de Abajo for providing the EPWE code and the financial support from the Spanish Ministry of Economy, Industry and Competitiveness (MINECO, Grant No. MAT2016-78293-C6 and FIS2016-75862-P), from the Basque Government (Grant No. IT-1255-19 and IT-756-13), from the regional Government of Aragon (RASMIA project), from the European Regional Development Fund (ERDF) under the program Interreg V-A Espa\~na-Francia-Andorra (Contract No. EFA
194/16 TNSI), from the SEV2015-0522, Fundacio' Privada Cellex and from the European Research Council (ERC-2012-StG 307760-SURFPRO).



\begin{thebibliography}{42}
\bibitem[{Foo()}]{Footnote1}
\bibinfo{journal}{The effect of a Co network on the Au Shockley state cannot be
  tested with ARPES, since at low coverages Co atoms aggregate forming clusters
  at herring-bone elbows (see SI)}  (????).
\end{thebibliography}
\begin{thebibliography}{99}
\bibitem[{Foo()}]{Footnote1}
\bibinfo{journal}{The effect of a Co network on the Au Shockley state cannot be
  tested with ARPES, since at low coverages Co atoms aggregate forming clusters
  at herringbone elbows (see SI)}.
\end{thebibliography}


\end{document}